# Detecting Sound Events Using Convolutional Macaron Net With Pseudo Strong Labels

Teck Kai Chan
*Faculty of Science, Agriculture and Engineering*
*Newcastle University Upon Tyne*
Singapore, Singapore
t.k.chan2@newcastle.ac.uk

Cheng Siong Chin
*Faculty of Science, Agriculture and Engineering*
*Newcastle University Upon Tyne*
Singapore, Singapore
cheng.chin@newcastle.ac.uk

*Abstract*— **In this paper, we propose addressing the lack of strongly labeled data by using pseudo strongly labeled data approximated using Convolutive Nonnegative Matrix Factorization. Using this set of data, we then train a novel architecture called the Convolutional Macaron Net (CMN), which combines Convolutional Neural Network (CNN) with MN, in a semi-supervised manner. Instead of training only a single model or using the Mean-teacher approach, we train two different CMNs synchronously using a curriculum consistency cost and a curriculum interpolated consistency cost. In the inference stage, one of the models will provide the frame-level prediction while the other model will provide the clip-level prediction. Our system outperforms the baseline system of the Detection and Classification of Acoustic Scenes and Events (DCASE) 2020 Challenge Task 4 by a margin of over 10% based on our proposed framework. By comparing with the top submission of the DCASE 2019 challenge, our system accuracy is also higher by 1.8%. On the other hand, as compared to the top submission of DCASE 2020, our accuracy is also marginally higher by 0.3%, even with fewer Transformer encoding layers. Our system remains robust on unseen YouTube evaluation dataset and has a winning margin of 0.6% and 6.3% against the top submission of DCASE 2019 and the baseline system.**

## I. Introduction

A Sound Event Detection (SED) system refers to a system capable of identifying the acoustic events captured in an audio clip as well as annotating the onset and offset of the identified event. Typically, a SED system developed using deep learning approaches can have much higher accuracy than a SED system developed using Hidden Markov Model or Gaussian Mixture Model [1]. However, deep learning approaches may require a large amount of strongly labeled data where the event tags and the corresponding onsets and offsets are known with certainty. The need for a large amount of strongly labeled data can be a limiting factor because such data is usually difficult and time-consuming to collect and thus limited to a small amount.

One possible solution is the use of weakly labeled data where only the event tags are known with certainty. Although crucial information about the acoustic event is missing, the analysis provided in [2] shown that proposals that utilized weakly labeled data can be promising in the development of an SED system. Some popular approaches include Multi-Instance Learning [3], [4], attention pooling [5], and semi-supervised learning using the mean-teacher approach [6].

Unlike the popular approaches, we propose to address the lack of strongly labeled data by using pseudo strongly labeled data. As shown in our earlier work [7], Nonnegative Matrix Factorization (NMF) [8] can be an effective pseudo labeling tool. In this paper, we improve in this aspect by proposing the use of Convolutive NMF (CNMF) [9] for pseudo labeling, which allows better spectral templates to be extracted and improves the quality of pseudo labeling. Using this set of data and synthetic strongly labeled and unlabeled data, we then train a novel architecture that combines Convolutional Neural Network (CNN) with a variant of Transformer [10], known as the Macaron Net (MN) [11] for SED. Such architecture will be referred to as the Convolutional Macaron Net (CMN) in the rest of the paper. The motivation for such a combination instead of the popular Convolutional Recurrent Neural Net (CRNN) comes from the fact that

1. The sequential nature of RNN can make it difficult for parallel computing [10], [12].
2. Transformer [10] was found to outperform RNN in various tasks such as language translation [10] and speech recognition
3. As seen in [12], detection accuracy using CNN with vanilla Transformer did not outperform CRNN. On the other hand, the combination of CNN with a variant of Transformer, known as the Conformer [14], proposed by Miyazaki et al. [15], achieved the best result in the DCASE 2020 Challenge Task 4. Such results show that Transformer variants can be better than vanilla Transformer.

We then propose using a new activation function known as Mish [16] to replace the conventional Rectified Linear Unit (ReLU) activation function in the entire architecture, which allows the detection accuracy to increase. Following our previous implementation [7], we propose training two different CMNs synchronously to pursue different targets wherein the inference stage, one of the models will provide the frame-level prediction. In contrast, the other model will provide the clip-level prediction. Such a framework is also related to [17], which utilized two CRNN for clip-level and frame-level prediction. However, [17] requires independent training of two models with different input types to achieve maximal results, while our implementation does not have such a prerequisite.

We then propose two new consistency costs: curriculum consistency cost and curriculum interpolated consistency cost. The main idea is to vary the confidence threshold so that consistency cost will not be calculated based on only highly confident predictions throughout the entire training process. This would in turn raise the detection accuracy of our system. Based on such a framework, our system can be competitive as compared to the other state-of-the-art, which we will show in the later section.

The rest of the paper is organized as follows. Section II describes the dataset used. Section III describes the proposed methodology, followed by the description of our experimental setup. Section V then presents the results and discussion. Finally, the paper ends with a conclusion.

## II. DATASET DESCRIPTION

In this paper, the DESED dataset [2] is used. This dataset is also used as the DCASE 2020 challenge task 4 data, which comprises 2595 strongly labeled synthetic audio clips, 1578 weakly labeled real audio clips, and 14412 unlabeled real audio clips. Each data subset consists of 10 event labels with a different distribution, representing the domestic environment. Each audio clip is 10s long and may contain more than one event. For audio clips containing multiple events, events may also overlap. In this paper, all three sets of data will be utilized. The model trained will be validated on the validation dataset, which consists of 1168 audio clips, and the public YouTube evaluation dataset, which consists of 692 audio clips.

## III. PROPOSED METHODOLOGY

### A. Data Preprocessing

In the preprocessing step, audio clips are resampled to 22050Hz. Audio clips not of 10s duration are either truncated or zero-padded to have a duration of 10s. A spectrogram is then tabulated for each clip using Short-Time Fast Fourier Transform with a window size of 2048 and a hop length of 345 and converted into mel spectrogram using 64 mel filterbanks. Such a setting would result in a mel spectrogram with 640 frames by 64 mel bins.

### B. Approximating Pseudo Strong Label Using CNMF

NMF [8] is a matrix decomposition method where the objective is to decompose a nonnegative matrix, $\mathbf{V} \in \mathbb{R}^{m \times n}$ into two nonnegative matrices, $\mathbf{W} \in \mathbb{R}^{m \times r}$ and $\mathbf{H} \in \mathbb{R}^{r \times n}$ where $r$ represents the number of components. In the SED domain, $\mathbf{W}$ represents the basis matrix while $\mathbf{H}$ represents the activation matrix. Thus, the temporal location of an event (i.e., frame-level label) can be found in a new clip by deriving the $\mathbf{H}$ of the clip using the extracted $\mathbf{W}$. While NMF is useful for analyzing data, it ignores the potential dependencies across successive columns of its input [9]. The fact that there is a sequence would not be apparent by examining the bases but would only be discovered by careful analysis of the basis weights [9]. To resolve this issue, CNMF is proposed, which extends the expression $\mathbf{V} \approx \mathbf{WH}$ as [9]

$$\mathbf{V} \approx \sum_{t=0}^{T-1} \mathbf{W}(t) \overset{t \rightarrow}{\mathbf{H}} \quad (1)$$

where $\rightarrow$ is the shift operator in the right direction, $t$ is the number of column shifts, and $T$ is the maximum allowable shift. $W$ and $H$ are updated as follows

$$\mathbf{W}(t) = \mathbf{W}(t) \otimes \left[ \left( (\mathbf{V}/\tilde{\mathbf{V}}) \cdot \left( \overset{t \rightarrow}{\mathbf{H}} \right)^T \right) / \left( 1 \cdot \left( \overset{t \rightarrow}{\mathbf{H}} \right)^T \right) \right] \quad (2)$$

$$\mathbf{H} = \mathbf{H} \otimes \left[ \left( \mathbf{W}(t)^T \cdot \left( \overset{\leftarrow t}{\mathbf{V}/\tilde{\mathbf{V}}} \right) \right) / \left( \mathbf{W}(t)^T \cdot 1 \right) \right] \quad (3)$$

Where $\otimes$ is the Hadamard product, $\leftarrow$ is the shift operator in the left direction, $\tilde{\mathbf{V}}$ is the approximated value of $\mathbf{V}$. Note that $\mathbf{H}$ is updated for each $t$. As shown in [9], such an extension can allow a better separation of audio mixtures. Therefore, it should also allow a better and more accurate basis matrix to be extracted, which improves the quality of pseudo labeling.

In this paper, we propose to extract the basis matrices from the synthetic audio clips to form different dictionaries for each event which results in ten different dictionaries. Since a clip may contain multiple events, frames that do not contain the event of interest are masked during the extraction of basis matrices. Using the dictionaries, we approximate a pseudo strong label for each weakly labeled clip. Each pseudo strong label has a size of 640 by 10 where the rows represent the event occurrence, and the columns represent the event label.

Based on the event label of the weakly labeled audio clip, we apply the corresponding dictionary to the mel spectrogram to derive $\mathbf{H}$. Values in $\mathbf{H}$ that are above 0.1 are considered as activated and converted to 1 and 0 otherwise. The augmented $\mathbf{H}$ will then replace the column in the pseudo strong label, which represents the event label. This process essentially allows us to obtain pseudo strong labels for all weakly labeled clips.

For better clarity, a synthetic audio clip which contains Speech from 0s to 5s is used as an example. A mel spectrogram is first tabulated and frames of the mel spectrogram are masked from 5s to 10s since they do not contain Speech. CNMF is subsequently applied to extract the spectral template. This process is applied to all synthetic audio clip containing Speech and the extracted spectral templates are consolidated to form a Speech dictionary.

Given a weakly labeled audio clip which contain Speech, we first tabulate the mel spectrogram. This mel spectrogram is then decompose using CNMF and the consolidated Speech dictionary. A threshold is then applied to the resulting activation matrix to find the activated frames. These activated frames will then act as the pseudo strong label for the Speech event. A similar procedure is performed for all weakly labeled clips that contain Speech.

In this paper, this set of data is termed as pseudo strongly labeled data. Once the pseudo labeling process is completed, all mel spectrograms are converted into log-scale using a logarithm operator, which will be used as our model input.

*C. Training CMN in a Semi-Supervised Framework*

Following our previous framework [7], we propose two different CMNs where one of the CMNs has no temporal compression, which will provide the frame-level prediction and will be named as Frame Level Model (FLM). The other CMN, which has higher temporal compression, will provide the audio tag and be called Clip Level Model (CLM). The model architecture for each CMN can be seen in Fig. 1. Besides the difference in temporal compression, the Clip Level model has more convolution layers with an increasing number of filters. For both models, we adopt a kernel size of 3 by 3, stride of 1 by 1, and padding of 1 by 1.

As seen in Fig. 1, output from the convolutional layers (referred to as Feature Map in Fig.2) will be pass to a Transformer encoding layer. As opposed to the vanilla Transformer encoding layer, we proposed using the MN encoding layer [11]. As seen in Fig. 2, the key difference between the two encoding layers lies in the number of position-wise feedforward module. As seen in Fig. 2, the MN encoder layer has an additional position-wise feedforward module before the multi-head attention module. Also, each position-wise feedforward module in the MN encoder layer is multiplied by 0.5. As explained in [11], by having such an arrangement, the stacked layers will be more accurate from the Ordinary Differential Equation's perspective and will leads to better performance in deep learning. In addition, Lu et al. [11] also demonstrated that Macaron Net could outperform Transformer on various supervised and unsupervised learning tasks.

Given that $\mathbf{x}_{FM}$ represents the feature map from the last convolutional layer and $\mathbf{y}_{FM}$ represents the output from the encoding layer. Mathematically, $\mathbf{y}_{FM}$ can be defined as

$$\mathbf{x'}_{FM} = LN\left(\mathbf{x}_{FM} + \frac{1}{2}PFF(\mathbf{x}_{FM})\right) \quad (4)$$

$$\tilde{\mathbf{x}}_{FM} = LN\left(\mathbf{x'}_{FM} + MHA(\mathbf{x'}_{FM})\right) \quad (5)$$

$$\mathbf{y}_{FM} = LN\left(\tilde{\mathbf{x}}_{FM} + \frac{1}{2}PFF(\tilde{\mathbf{x}}_{FM})\right) \quad (6)$$

Where LN represents Layer Normalization. PFF represents the Position-wise FeedForward module which contain the following operation.

$$PFF(\mathbf{x}_{FM}) = \mathbf{W}_2\left(f\left(\mathbf{W}_1\mathbf{x}_{FM} + \mathbf{b}_1\right)\right) + \mathbf{b}_2 \quad (7)$$

Where $\mathbf{W}_1$, $\mathbf{B}_1$ and $\mathbf{W}_2$, $\mathbf{B}_2$ represents the parameters of the first and second fully connect layer. $f(.)$ represents an activation function.

MHA represents the Multi-Head Attention module and contains the following operations.

$$MHA(\mathbf{Q},\mathbf{K},\mathbf{V}) = Concat(\mathbf{H}_1,\cdots,\mathbf{H}_h)\mathbf{W}_o \quad (8)$$

$$\mathbf{H}_i = Attention\left(\mathbf{Q}\mathbf{W}_i^Q, \mathbf{K}\mathbf{W}_i^K, \mathbf{V}\mathbf{W}_i^V\right) \quad (9)$$

$$Attention(\mathbf{Q},\mathbf{K},\mathbf{V}) = softmax\left(\frac{\mathbf{Q}\mathbf{K}^T}{\sqrt{d_k}}\right)\mathbf{V} \quad (10)$$

Where $\mathbf{Q}$, $\mathbf{K}$ and $\mathbf{V}$ represents the query key and value and is equivalent to the output from the first PFF module. $\mathbf{W}_i^Q$, $\mathbf{W}_i^K$, $\mathbf{W}_i^V$ are the learnable parameters for the i head. $\mathbf{W}_o$ is the final linear parameter matrix applied on the concatenated feature vector. $d_k$ represents the dimension of the key.

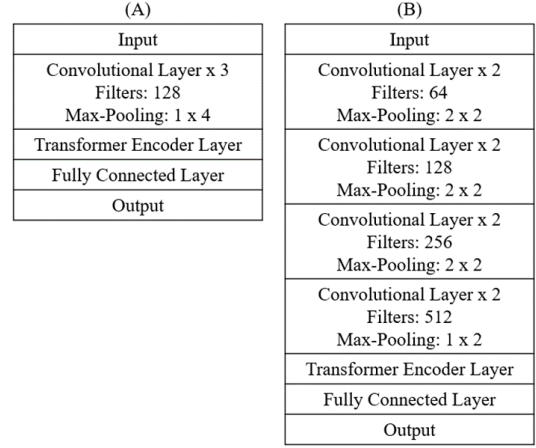

Fig. 1. (A) FLM (B) CLM

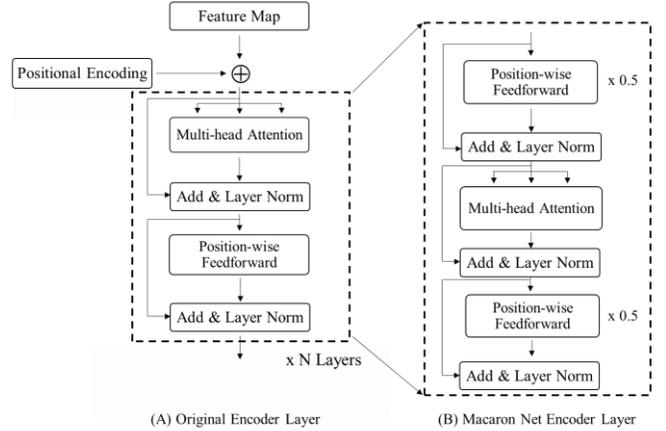

Fig. 2. Difference between the encoder layers

In this paper, the number of units in the feedforward module is set to the same value as the filter size of the last convolutional layer of each model.

As opposed to the use of ReLU, we propose the use of Mish [16] as the activation function after each convolution operation and in the position-wise feedforward module. Unlike ReLU, Mish is continuously differentiable which is a preferable characteristic because it avoids singularities and undesired side effects when performing gradient-based optimization [16]. Moreover, Mish was shown to be better in terms of performance and stability as compared to the other activations functions across different image-related tasks [16]. Thus, we hypothesized that that Mish can also improve the accuracy of SED. Given that $x$ represents the input to the activation function (usually the output from a convolutional or fully connect layer), Mish can be defined as [17]

$$f(x) = x\tanh(\text{softplus}(x)) \quad (11)$$

Where

$$\text{softplus}(\mathbf{x}) = \ln(1+e^{\mathbf{x}}) \quad (12)$$

In this framework, the two models are trained synchronously, where FLM is forced to learn from the CLM. As shown in our earlier work [7], by enforcing the prediction of FLM to be close to CLM, the frame-level prediction can be more accurate. In this framework, four different loss components are introduced. The first loss component can be defined as the Binary Cross Entropy (BCE) loss between the frame-level prediction by FLM, $\mathbf{FLM}_F$, and the frame-level ground truth (i.e., pseudo strong labels and/or synthetic strong labels), $\mathbf{Y}_F$, of an audio clip and is given as

$$L_F = \text{BCE}(\mathbf{FLM}_F, \mathbf{Y}_F) \quad (13)$$

The second loss component is then defined as the BCE loss between the clip level prediction by CLM, $\mathbf{CLM}_C$, and the audio tag, $\mathbf{Y}_C$, of an audio clip and is given as

$$L_C = \text{BCE}(\mathbf{CLM}_C, \mathbf{Y}_C) \quad (14)$$

The third loss component is defined as the consistency loss between the clip level prediction given by FLM and CLM. Since the prediction by FLM is in frame level, we apply a temporal max pooling on $\mathbf{FLM}_F$ to obtain the clip level prediction $\mathbf{FLM}_C$. Rather than the use of BCE, we propose the use of Mean Square Error (MSE) as it was found to be a better consistency loss in [18]. It was shown that the deep learning model begins learning the easier example before moving to a harder example [19]. As such, we propose a curriculum consistency cost between $\mathbf{FLM}_C$ and $\mathbf{CLM}_C$, such that FLM will only enforce its prediction to be consistent with CLM provided if CLM is confident with its prediction at each learning stage. This is given as

$$L_{\text{con}} = \begin{cases} \text{MSE}(\mathbf{FLM}_C, \mathbf{CLM}_C), \max(\mathbf{CLM}_C) > \lambda_{\text{curr}} \\ 0, \text{otherwise} \end{cases} \quad (15)$$

where $\lambda_{curr}$ is the current confidence level based on the training progression and is defined as

$$\lambda_{\text{curr}} = \lambda_{\min} + 0.5(\lambda_{\max} - \lambda_{\min})(1 + \cos(\pi T_{\text{curr}} / T_i)) \quad (16)$$

where $\lambda_{\max}$ and $\lambda_{\min}$ is the maximum and minimum confidence level and is set at 0.9 and 0.6. Thus, $\lambda_{curr}$ will slowly decrease along a cosine curve from 0.9 to 0.6.

As suggested in [7], if CLM has yet to achieve optimal detection capability (i.e., low audio tagging accuracy at the early training stage), FLM may be force to produce incorrect predictions at the early stage if $\lambda_{curr}$ is set as a low value which can lead to performance degradation. However, setting a $\lambda_{curr}$ throughout the training phase may restrict the models to learn from only easier training examples. Thus, by varying the confidence level, the models will consider the high confidence (lower difficulty) examples first before the low confidence (higher difficulty) examples. This avoid the scenario that FLM has to learn from CLM when CLM has yet to achieve optimal detection accuracy and at the same time allows the models to learn from more training examples.

The fourth loss component is the interpolated consistency cost [21] between the two models' prediction on the unlabeled data. Based on the concept of mixup [22], interpolated consistency tries to enforce the $\mathbf{FLM}_{UC}$ to be similar to $\mathbf{CLM}_{UC}$ where $\mathbf{FLM}_{UC}$ and $\mathbf{CLM}_{UC}$ represents the prediction by FLM and CLM on an unlabeled sample which is given as

$$\mathbf{FLM}_{UC} = M(\text{FLM}(\text{mixup}(\mathbf{u}_1, \mathbf{u}_2))) \quad (17)$$

$$\mathbf{CLM}_{UC} = \text{mixup}(\text{CLM}(\mathbf{u}_1), \text{CLM}(\mathbf{u}_2)) \quad (18)$$

where $M(.)$ is the temporal max pooling operator, $u_1$ and $u_2$ represents the unlabeled sample 1 and 2. Similarly, the loss is only calculated if CLM is confident with its prediction at the different learning stages. As an additional measure to prevent a suboptimal solution, a weighing parameter, $w$, is included to regularize the contribution of this loss. Thus, this is given as

$$L_{\text{inter}} = \begin{cases} w \times MSE(\mathbf{FLM}_{UC}, \mathbf{CLM}_{UC}), \max(\mathbf{CLM}_{UC}) > \lambda_{\text{curr}} \\ 0, \text{otherwise} \end{cases} \quad (19)$$

where $w$ is defined as [18]

$$w = \exp(-5(1 - T_{\text{curr}} / T_i)) \quad (20)$$

The training of the model and hyperparameter analysis are presented in the next few sections.

## IV. EXPERIMENTAL SETUP

In this paper. The models are trained in 2 different phases; 1) warm-up phase and 2) model-tuning phase. In the first phase, models are trained using only synthetic and pseudo strongly labeled data where they are augmented with Gaussian noise, time mask, and frequency mask. A batch size of 32 is used and are even split between the two data types. As this is the warm up phase, models are trained with an increasing Learning Rate (LR) and the LR at each iteration is defined as [20]

$$LR_{\text{curr}} = LR_{\min} + 0.5(LR_{\max} - LR_{\min})(1 + \cos(\pi(T_i - T_{\text{curr}})/T_i))$$
$$(21)$$

where $LR_{\text{curr}}$ is the current LR. $LR_{\max}$ and $LR_{\min}$ are the maximum and minimum learning rate set as 0.0014 and 1e-6 respectively. $T_i$ is the total iterations during the warm up phase, which we set as the total number of iterations in 10 epochs. As unlabeled data is not utilized in this phase, the total loss is the summation of $L_F$, $L_C$ and $L_{\text{con}}$. In this phase, curriculum consistency cost is not utilized and $\lambda_{\text{curr}}$ which is used in the calculation of $L_{\text{con}}$ is set as 0.9. Based on the losses calculated, the models are updated using Lookahead [23] with an alpha of 0.5 and a step size of 20 together with Adam [24].

In the model tuning phase, models are trained with synthetic, pseudo strongly labeled and unlabeled data. Augmentation techniques remains the same for synthetic and pseudo strongly labeled but unlabeled data is only augmented with Gaussian noise. a batch size of 64 is used where half of them is a mixture of synthetic and pseudo strongly labeled data, and the other half is the unlabeled data. LR at each iteration is defined similar to equation (16) but with different bounds where $LR_{max}$ and $LR_{min}$ are set as 0.0014 and 1e-6 respectively. $T_i$ is set as the total number of iterations in 100 epochs. Since all types of data are used, the total loss in this phase is the summation of $L_F$, $L_C$, $L_{con}$ and $L_{inter}$. In this phase, $\lambda_{curr}$ is calculated according to equation (16). Based on the losses calculated, the models are updated using Lookahead [23] with an alpha of 0.5 and a step size of 20 together with Adam [24].

In the inference stage, CLM is used to predict the occurrence of any event in an audio clip. A sound event is considered to be present if the predicted probability by CLM is larger than 0.5. Based on the audio tag, the temporal location of the event is then determined using the corresponding output from FLM. To improve the frame-level prediction accuracy, outputs from FLM are first smoothed using a median filter. Events that are shorter than 0.1 seconds are removed, and similar events are concatenated if the difference between the offset and onset is lesser than 0.2. In this paper, the accuracy of our system is measured using the event-based metric [25].

## V. RESULTS AND DISCUSSION

We began our experiments using one MN encoder layer with four attention heads. We first investigated the importance of warm-up which is considered a critical component to train a Transformer [10]. As seen in Table I, models can benefit from the use of warm-up which brings a maximum accuracy increment of 2.4%.

We then varied the number of attention heads and layers of the encoder layer. As seen in Table II, using a single layer with four heads is sufficient, and any more can degrade the accuracy.

An ablation study was performed to investigate the importance of $L_{inter}$ and $L_{con}$. As seen in Table III, the two losses are critical, and ablating them would result in an accuracy drop, although the impact of ablating $L_{inter}$ is lower due to the regularizing term, $w$. Such results coincide with our previous study in [7] where consistency losses are shown to be critical components.

We then compare the use of curriculum consistency losses against consistency losses with a constant confidence threshold of 0.9. As seen in Table IV, our proposed curriculum consistency losses also perform better than consistency losses with a constant confidence threshold by 1.4%. This shows the importance of including not just the confident prediction during the training process.

Since curriculum consistency losses are shown to be better than the use of constant threshold, we investigate on the optimal $\lambda_{min}$ value. As seen in Table V, while it is important to include lesser confident examples in the losses calculation but it can have a negative impact if $\lambda_{min}$ is set lower than 0.6. On the other hand if $\lambda_{min}$ is set too high (i.e., 0.8), it does not bring any benefits.

We then investigate on the effect of using different alpha and step size for Lookahead [23]. Zhang et al [23] explained that Lookahead reduces the variance and is less sensitive to suboptimal hyperparameters and can reduce the need for extensive hyperparameter tuning. However, as shown in Table VI, we find that there is still a need to tune alpha and step size. As shown in Table VI, while Lookahead does provide accuracy improvent but improvement will only be marginal if suboptimal parameters are used.

We then compared the effectiveness of CNMF against NMF as a pseudo labeling tool, and the results shown in Table VII indicate that CNMF is a better approximator. However, this improvement comes at the expense of computational time due to the different shift operations. We then compared against two other types of pseudo labels. 1) A system trained with pseudo strong labels where all frames are annotated as positive (i.e., containing an event) and 2) a system trained with pseudo strong labels where all frames are annotated as negative (i.e., does not contain any event). Results shown in VII indicate that our proposed labeling method using either NMF or CNMF can be much better than the naïve assumption where all frames are either annotated as positive or negative.

Subsequently, we investigated the importance of the positional encoding module, which is used to inject information about the relative position of an input sequence. As seen in Table VIII, there is not much difference in accuracy if the positional encoding module is ablated. This may suggest either there is no use for PE in the SED application or the use of sine and cosine functions of different frequencies to inject sequence information may not be effective.

We then compared the use of ReLu against Mish [16], and the results in Table VIII show that Mish does outperform ReLu. Finally, we tested the system using the vanilla Transformer encoder layer. The result shows that the vanilla Transformer encoder layer performs poorly against the system using the MN encoder layer.

We then compared against the non-ensembled systems from the top submission in DCASE 2019, DCASE 2020 Challenge task 4, and the baseline system. As seen in Table V, our system outperforms the baseline by a margin of over 10% and also the 1st place submission in DCASE 2019 by 1.8%. At the same time, we also show that using a single layer four head MN encoding layer is sufficient to produce comparable results as the 1st place submission in DCASE 2020, which utilized a four layers four heads Conformer encoding layer. By applying our system on unseen public YouTube evaluation dataset (Note that this dataset is not the challenge evaluation dataset which is unavailable to the public), accuracy only has a 0.2% deviation, which shows its robustness. Our system also outperforms the baseline system by 6.3% and has a narrow margin against the 1st place submission of DCASE 2019 by 0.6%.

## VI. CONCLUSION

In this paper, we proposed a novel framework with several improvements to train a SED system. We showed that the use of pseudo labels can be a viable solution to the lack of strongly labeled data problem and results also indicates that CNMF can be an effective solution for pseudo labeling. Results also shows that our proposal of integrating curriculum learning and consistency losses can bring an accuracy improvement to our models. Based on our results, our system can provide competitive accuracy to the other state of the art. However, there is still a large room for improvement. As our future work, we plan to further investigate the effectiveness of combining CNN and Transformer.

Table. I. System accuracy with different warm-up epochs

| Warm-up Epochs | Event-Based F1-Score |
|---|---|
| 0 | 43.9 |
| 5 | 44.4 |
| 10 | **46.3** |
| 20 | 44.4 |

Table. II. System accuracy with different encoding layers and heads

| | | Layer | | | | | |
|---|---|---|---|---|---|---|---|
| | | 1 | 2 | 3 | 4 | 5 | 6 |
| Head | 4 | **46.3** | 45.2 | 44.8 | 42.3 | 43.6 | 44.2 |
| | 8 | 45.1 | 44.9 | 43.2 | 42.5 | 41.9 | 42.1 |
| | 16 | 45.4 | 44.5 | 44.9 | 43.0 | 42.8 | 44.0 |
| | 32 | 45.1 | 44.1 | 44.3 | 42.9 | 43.3 | 42.9 |

Table. III. Ablation of $L_{inter}$ and $L_{con}$

| | Event-Based F1-Score |
|---|---|
| Include $L_{inter}$ and $L_{con}$ | **46.3** |
| Ablate $L_{inter}$ | 45.8 |
| Ablate $L_{inter}$ and $L_{con}$ | 44.9 |

Table. IV. Comparison of accuracy with and without curriculum consistency losses

| Curriculum consistency losses | Constant $\lambda_{curr}$ of 0.9 |
|---|---|
| **46.3** | 44.9 |

Table. V. Effect of $\lambda_{min}$ on accuracy

| $\lambda_{min}$ | Event-Based F1-Score |
|---|---|
| 0 | 43.6 |
| 0.2 | 45.2 |
| 0.4 | 45.3 |
| 0.6 | **46.3** |
| 0.8 | 44.3 |

Table. VI. Parameter analysis for Lookahead

| | | Step size | | | |
|---|---|---|---|---|---|
| | | 5 | 10 | 20 | Ablate Lookahead |
| Alpha | 0.1 | 44.7 | 45.6 | 45.5 | 44.8 |
| | 0.5 | 45.5 | 45.2 | **46.3** | |
| | 1 | 44.9 | 45.3 | 45.8 | |

Table. VII. Accuracy of system trained with different pseudo labels

| Setting | Event-Based F1-Score |
|---|---|
| Pseudo labeling using CNMF | **46.3** |
| Pseudo labeling using NMF | 45.0 |
| All frames considered to contain event | 36.0 |
| All frames are unlabeled | 35.1 |

Table. VIII. Architecture accuracy using different settings

| Setting | Event-Based F1-Score |
|---|---|
| Proposed | **46.3** |
| Ablate positional encoding | 46.2 |
| Using ReLu | 43.6 |
| Vanilla Transformer | 42.5 |

Table. IX. Comparison against the other systems

| Methodology | Event-Based F1-Score | |
|---|---|---|
| | Validation | YouTube |
| Proposed | **46.3** | 46.1 |
| (1st in DCASE 2020) CNN + Conformer [15] | 46.0 | - |
| (1st in DCASE 2019) CNN [27] | 44.5 | 45.5 |
| Baseline [26] | 35.6 | 39.8 |